\documentclass[preprint,5p,times,twocolumn]{elsarticle}

\usepackage{epsfig}
\usepackage{booktabs}
\usepackage{amsmath, amsthm, amsfonts}
\usepackage{xcolor}
\usepackage{float}
\usepackage{dsfont}

\usepackage{amssymb}

\usepackage{upgreek}
\journal{xxxx}

\begin{document}

\begin{frontmatter}



\title{Synthesizing Affective Neurophysiological Signals Using Generative Models: A Review Paper}

\author[label2]{Alireza F. Nia}
\author[label2]{Vanessa Tang}
\author[label2]{Gonzalo Maso Talou}
\author[label2]{Mark Billinghurst}

\affiliation[label2]{organization={Auckland Bioenginering Institute},
            addressline={70 Symonds Street},
            city={Auckland},
            postcode={1010},
            country={New Zealand}}

\begin{abstract}
The integration of emotional intelligence in machines is an important step in advancing human-computer interaction. This demands the development of reliable end-to-end emotion recognition systems. However, the scarcity of public affective datasets presents a challenge. In this literature review, we emphasize the use of generative models to address this issue in neurophysiological signals, particularly Electroencephalogram (EEG) and Functional Near-Infrared Spectroscopy (fNIRS). We provide a comprehensive analysis of different generative models used in the field, examining their input formulation, deployment strategies, and methodologies for evaluating the quality of synthesized data. This review serves as a comprehensive overview, offering insights into the advantages, challenges, and promising future directions in the application of generative models in emotion recognition systems. Through this review, we aim to facilitate the progression of neurophysiological data augmentation, thereby supporting the development of more efficient and reliable emotion recognition systems. 
 
\end{abstract}




\end{frontmatter}


\section{Introduction}\label{sec:intro}

Emotions play a pivotal role in shaping human experiences, interactions, and decision-making processes~\cite{eMarg1995}. The capacity for computers to recognize and interpret human emotions is an essential factor in enhancing human-computer interaction (HCI) and fostering seamless communication between humans and machines. The recent advancements in deep learning and artificial intelligence (AI), coupled with the proliferation of lightweight, multimodal wearable devices that can record physiological data, have spurred a surge in research efforts aimed at developing end-to-end emotion recognition systems capable of recognizing a wide range of emotions\cite{mMaithri2022}. These systems typically leverage various modalities, such as facial expressions~\cite{fCanal2022}, speech~\cite{ySingh2022}, physiological signals~\cite{yWang2022}, and behavioral cues~\cite{iDing2022}, to understand emotions. Among these modalities, neurophysiological data, specifically electroencephalography (EEG) and functional near-infrared spectroscopy (fNIRS) have gained increasing attention due to their consistency across cultures and genders\cite{zLiu2021}, robustness to the constraints of external indicators such as the possibility of hiding or feigning\cite{sWioleta2013,yWang2022}. Along with recent technological advances that have enabled easy data acquisition and low-cost implementation, many neurophysiological emotion recognition systems have been introduced that are able to accurately identify a wide range of types and intensity of emotions\cite{mBalconi2015,xLi2022}.

However, despite the effectiveness of the neuronal data in differentiating between different mental states, these systems often suffer from a lack of generalizability to new, unseen data and have limited ability to perform cross-subject validation. The challenges are largely attributed to the issue of data scarcity that arises from small sample sizes in neurophysiological signals, which can lead to potential overfitting and reduced statistical power\cite{yLuo2020}. This issue is even more pronounced when building end-to-end systems that leverage deep learning models, as the size of the training dataset is a critical determinant of the system's configuration and performance. A sufficiently large sample size is necessary to improve system performance, enhance the system's ability to generalize to unseen data, and reduce the possibility of model bias by directly influencing the bias-variance trade-off\cite{rDwivedi2020}. The complexity of experimental designs, which can involve sophisticated equipment or complex tasks, has made HCI experiments time-consuming and uncomfortable for participants and led to a high attrition rate in these studies. Consequently, the publicly accessible multimodal affective databases, which include neuronal data primarily in the form of EEG signals (e.g., SEED\cite{wZheng2015}, DEAP\cite{sKoelstra2011}, DREAMER\cite{DREAMER}, and MAHNOB-HCI\cite{jLichtenauer2011}), is limited in terms of size and scope. They typically contain data from fewer than 40 subjects and 50 trials on average. In some cases, the data is imbalanced, with a significant skew toward specific labels. Additionally, the variability in experimental paradigms and measured biomarkers has posed challenges in combining data across studies and effectively utilizing these existing datasets.

In recent years, researchers have employed different techniques in a variety of neurophysiological emotion recognition systems to address data scarcity~\cite{eLashgari2020}.  These techniques can be divided into two main categories: model-focused and data-focused. 

In the model-focused approach, different deep learning models have been employed for multi-source fusion~\cite{zLan2018}, mitigation of the adverse effect of subject/paradigm differences~\cite{zWan2021, wLi2021}, and reduction of the impact of small sample size on model performance using different regularization techniques~\cite{xShi2004}. One promising method in this category is domain adaptation which focuses on leveraging knowledge acquired from a source domain (with abundant labeled data) to improve the learning performance in a target domain (with limited or no labeled data), despite potential differences between the two domains. The goal is to adapt models or algorithms trained on the source domain to generalize effectively to the target domain, mitigating the effects of domain shift or distribution discrepancies that may arise due to differences in data distribution, feature space, or task-specific characteristics. Researchers have utilized this technique for cross-subject EEG-based emotion classification based on a small set of labeled data~\cite{zLan2018-domain,gBao2021-domain}. Other methods in this category include transfer learning and the use of pre-trained models~\cite{jLi2019,sKhare2020}. While all the techniques are effective in addressing the data scarcity problem, they suffer from potentially requiring large amounts of computational resources for fine-tuning, availability of source domain or biased source domain, and dissimilarity between source and target domain. 

On the other hand, data-focused approaches focus on increasing the number of samples with the aim of improving the accuracy and robustness of the model\cite{gZhang2022} via two subcategories of methods: deterministic and non-deterministic data augmentation methods. Deterministic data augmentation techniques involve applying predefined transformations to the existing data, either in data or feature space. These techniques are deterministic because the outcome is directly determined by the input and transformation applied. The success of this method in computer vision has led researchers to apply it in other domains, such as speech recognition, natural language processing, audio analysis, medical image processing, and affective computing. The two main conventional techniques in this category are geometric transformation and noise addition. Given the applicability to time series data, the latter has been used more extensively in neuronal data. Other than noise addition\cite{gZhang2022,Tunnel2022}, many sophisticated augmentation techniques in this category have been employed in neurophysiological affective systems, including Mixup\cite{hZhang2017mixup}, Up-sampling\cite{UpsamplingLuo2022, Chen2021}, empirical component-based~\cite{EMDSingh2023} and independent component-based methods~\cite{ICAKang2022}, and others~\cite{Ghosh2020}. The popularity of these methods lies in three main factors, including ease of implementation, computation efficiency, and, more importantly, interpretability of the outcome. Besides the versatility of these techniques for different data types, the non-stationarity of neurophysiological signals casts specific limitations when it comes to augmenting these data. The simplicity of these techniques might fail to introduce sufficient variability to the existing dataset for model generalization and consequently increase the possibility of overfitting because of the huge similarity between generated samples with the original dataset. In addition,  predefined transformations may not successfully capture the individual differences and fail to represent inter-subject variability in the generated samples. Finally, applying deterministic methods is a manual process and requires careful selection of appropriate transformations by researchers, which can be challenging to identify.  

Non-deterministic techniques are another branch of data-focused approaches that involve training deep learning models, particularly generative models, to learn the underlying data distribution and generate new samples by sampling from the learned distribution. These techniques are non-deterministic, as the generated samples are random instances drawn from the learned distribution. Recent advancements in generative models such as generative adversarial networks (GANs)~\cite{iGoodfellow2020}, and variations auto-encoders (VAEs)~\cite{dKingma2013} proved successful in computer vision for generating realistic images and videos, resulting in an increase in popularity across different areas of research~\cite{lYang2022}. These methods generate more diverse samples by learning the underlying data distribution and capturing the complexity of human emotions can overcome the shortcomings of the deterministic methods in handling neurophysiological signals in affective systems. In this paper, we present a review of the existing literature on synthesizing neurophysiological data using non-deterministic methods for emotion recognition systems. By providing a summary of employed techniques and describing their pros and cons, this paper seeks to inspire future research on the data scarcity problems in neurophysiological emotion recognition systems.

The remainder of the paper is organized as follows. In Section \ref{sec:background}, background knowledge about generative models is provided. Section \ref{sec:method} outlines the methodology for this review and the papers that meet the criteria are described in Section \ref{sec:approaches}. Section \ref{sec:discussion} summarises the trend observed in utilising generative models, and Section~\ref{sec:conclusion} concludes the paper and identifies future research direction. 

\section{Scientific Background} \label{sec:background}
Non-deterministic techniques involve training deep learning models to learn the underlying data distribution and generate new samples by sampling from the learned distribution. These techniques are probabilistic in nature, as the generated samples are random instances drawn from the learned distribution. Examples include GANs, VAEs, restricted Boltzmann machines (RBMs), deep belief networks (DBNs), and autoregressive models. Affective computing researchers have only started using GANs and VAEs, or a hybrid approach of both; hence, in the following subsections, we describe the mathematical background of GAN and VAEs. 

\subsection{Autoencoder}
Autoencoder (AE), introduced in late 1980~\cite{hBourlard1988}, is a nonlinear dimensionality reduction technique that consists of two neural networks working in tandem, an encoder $f_{\theta_{e}}$ and a decoder $g_{\phi_{d}}$,

\begin{align}
    f_{\theta_{e}}(x) &= Net(x, \mathbf{W}_e, \mathbf{b}_e) = z\\
    g_{\phi_{d}}(z) &= Net(z, \mathbf{W}_d, \mathbf{b}_d) = \hat{x} \\
    Net(x,\mathbf{W},\mathbf{b}) &= l_{L}(x_L,W_L,b_L) \circ \cdots \circ l_1(x,W_1,b_1) \\
    l(x_i,W_i,b_i) &= a(W_i x_i + b_i) = x_{i+1}
\end{align}
where $x \in \mathds{R}^{N}$ is the input data, $z \in \mathds{R}^{M} (M<N)$ is the latent representation, $\hat{x} \in \mathds{R}^{N}$ is the reconstructed data, $\mathbf{W}_e=[W_1,\ldots,W_{L_e}]$ and $\mathbf{W}_d=[W_1,\ldots,W_{L_d}]$ are the encoder and decoder weights, $\mathbf{b}_e=[b_1,\ldots,b_{L_e}]$ and $\mathbf{b}_d=[b_1,\ldots,b_{L_d}]$ are the encoder and decoder biases for the networks $Net$ with layers $l_i$ and activation functions $a$. The encoder $f$ and decoder $g$ functions are parameterized by $\theta_{e}$ and $\phi_{d}$, respectively. The encoder $f_{\theta_{e}}(x)$ aims to capture the hidden patterns in input data $x$ and compresses them into its constructing components called latent representation/bottleneck $z$. On the other hand, the decoder $g_{\phi_{d}}$ attempts to reconstruct the input data from its latent representation. The objective of this autoencoder is to reconstruct its input as accurately as possible, which is achieved through minimizing a cost function called \textit{reconstruction loss}, defined as

\begin{align}
    \mathcal{L}(x, \hat{x}) = \|x-\hat{x}\|^2.
\end{align}

\begin{figure}
    \centering
    \includegraphics[scale=0.15]{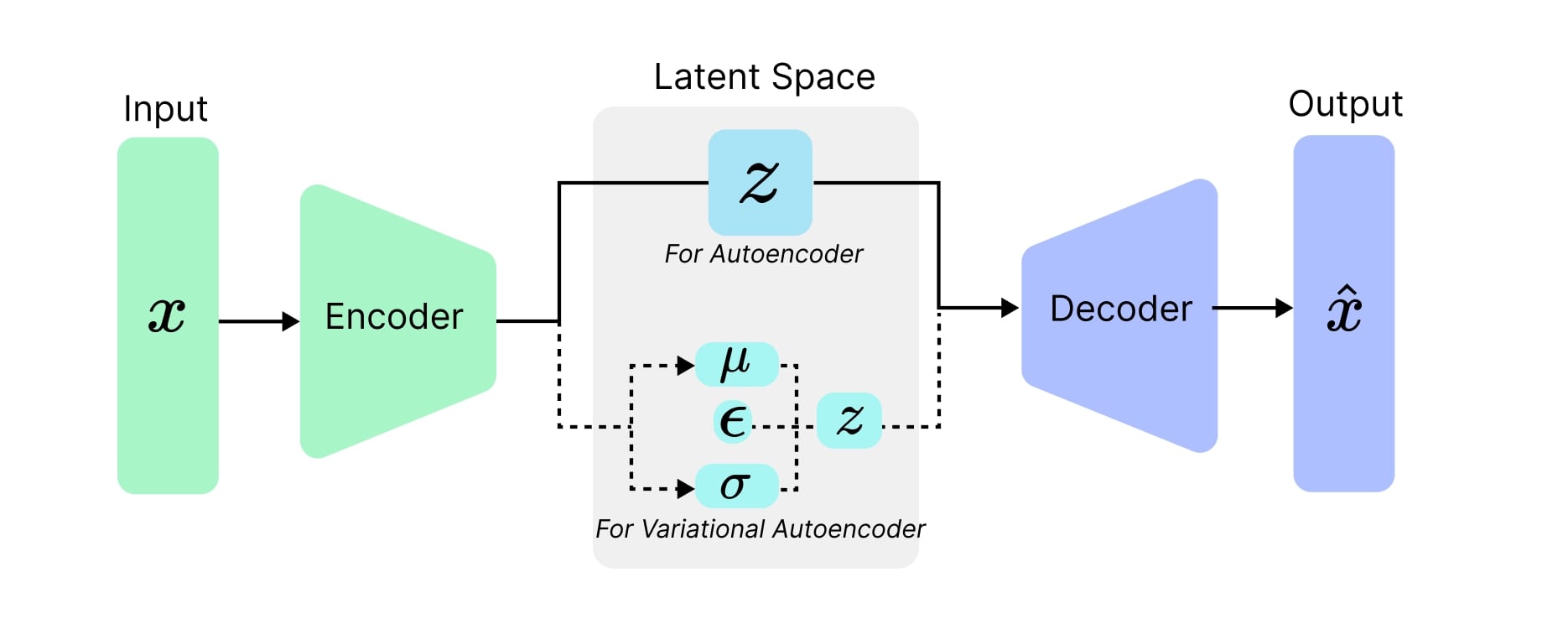}
    \caption{General Schematic of constructing components of AE and VAE.}
    \label{fig:AE}
\end{figure}
As a representation learning method, AE builds a latent representation with lower dimensionality than input and with less correlated/redundant features of input data~\cite{aMechelli2020}. 

There are various types of AE, each with different objectives, such as denoising AE, sparse AE, and variational autoencoders (VAEs). VAEs have gained popularity for data augmentation in comparison to other variants due to its probabilistic nature and ability to model the underlying data distribution. 

VAEs, developed by Kingma et al.\cite{dKingma2013}, in contrast to a simple autoencoder, model the latent space as a probabilistic distribution. The encoder in VAE, called \textit{inference}, is a probability function $q_{\phi}(z|x)$ that maps the input data to the latent space, which can be defined by parameters of a distribution, typically a Gaussian distribution $z \sim N(\mu, \sigma^2)$. The decoder in VAE is a generative model $p_{\theta}(x|z)$. The decoder reconstructs the input data from a sample drawn $z$ from the estimated latent distribution. The objective of the VAE, other than better reconstruction of the input data from latent space, is also to have a better approximation of latent prior, usually defined as $p(z) \sim \mathcal{N}(0, I)$, through variational posterior distribution $q_{\phi}(z|x)$. To achieve this, VAE is trained to maximize a derived lower bound for the likelihood of the data, also called evidence lower bound (\textit{ELBO}) to optimize both the reconstruction error and the Kullback-Leibler (KL) divergence between the latent distribution $q_{\phi}(z|x)$ and a prior distribution $p(z)$. This function can be rewritten as a loss function (i.e., negated ELBO) as follows

\begin{align}
    \mathcal{L}_{VAE}(\theta, \phi) = \begin{aligned}[t] &-\mathbb{E}_{z \sim q_{\theta}(z|x_{i})}[\log_{p_{\phi}}(x_{i}|z)] \\
    &+ \mathbb{KL}(q_{\theta}(z|x_{i}) || p(z) ).
    \end{aligned}
\end{align}
In general, Variational Autoencoders (VAEs) demonstrate improved generative capabilities, greater resilience against overfitting in comparison to basic Autoencoders (AEs), and enhanced representation learning. As a result, VAEs prove to be a suitable technique for generating diverse and realistic samples for neuronal emotion recogntiion systems.

\subsection{Generative Adversarial Networks (GANs)}
The Generative Adversarial Network (GAN), also referred to as vanilla GAN, was introduced by Ian Goodfellow et al. in 2014~\cite{iGoodfellow2020}. This pioneering generative model has since transformed synthetic data generation methods, particularly in the realm of computer vision. Unlike VAEs that utilize an encoder and decoder network to infer the data distribution and generate new data points by sampling from that distribution, GANs employ a distinct architecture consisting of a \textit{generator} and a \textit{discriminator} within an \textit{adversarial} framework to synthesize new samples. 
\begin{figure}
    \centering
    \includegraphics[scale=0.15]{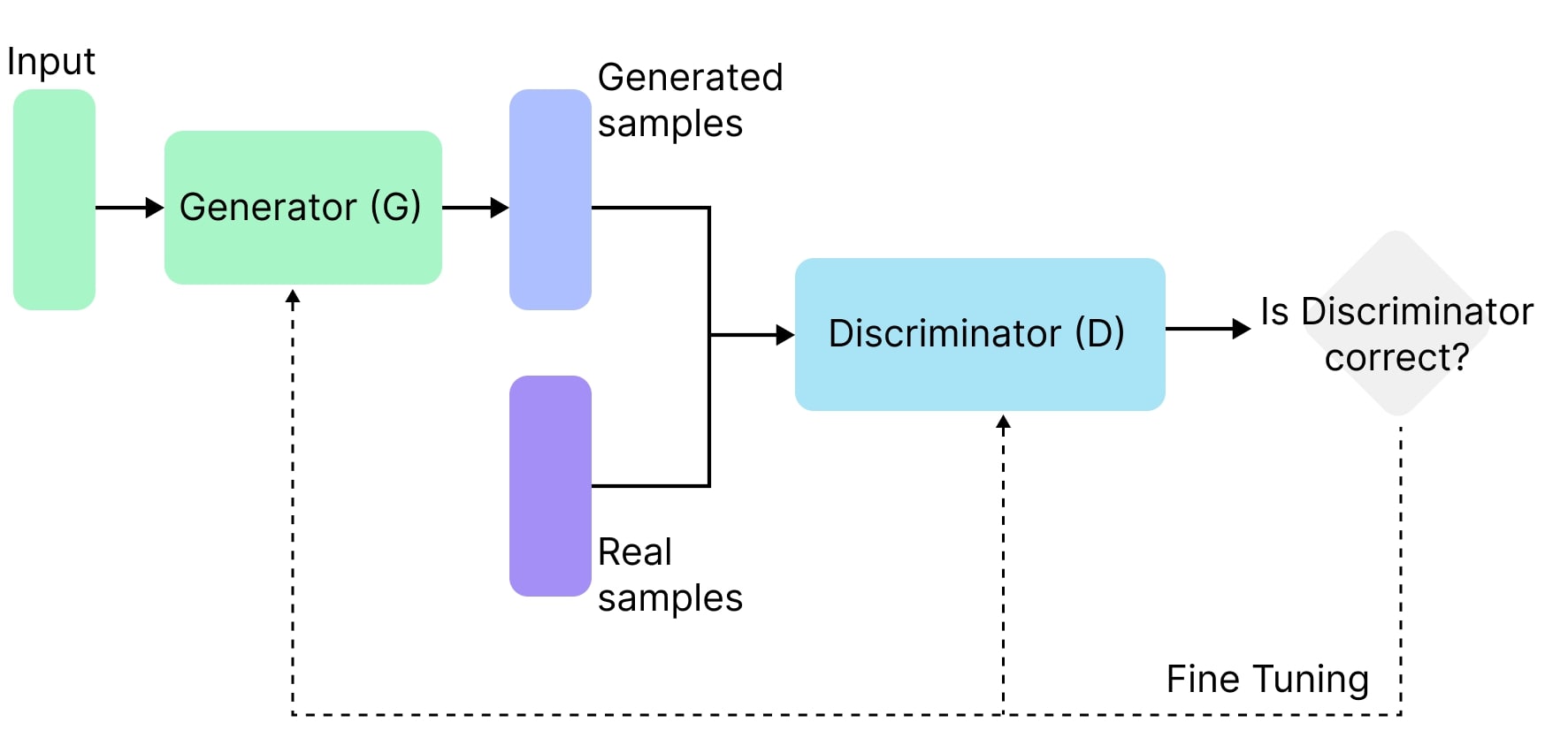}
    \caption{Building components of the GANs and the flow of data}
    \label{fig:GAN}
\end{figure}
Within this adversarial framework, the generator network $G$ captures the data distribution and generates new samples, while the discriminator network $D$ estimates the probability of a sample originating from the training data rather than from the generator. Mathematically, the generator's objective is to implicitly define a probability distribution $p_g$ that transforms the prior noise distribution $p_z(z)$ into a distribution resembling the real data distribution $p_{data}$. Conversely, the discriminator functions as a classifier, producing a single scalar $D(s)$ representing the probability of a sample $s$ being real ($s \sim p_{data}$) rather than generated ($s \sim p_g$). In this adversarial framework, the generator aims to minimize $log(1-D(G(z)))$ by generating samples $G_{z \sim p(z)}(z)$ similar to real data, while the discriminator seeks to maximize its accuracy in distinguishing real samples from generated ones by maximizing $\log(D_{x \sim p_{data}}(x)) + \log(1 - D_{z \sim p(z)}(G(z)))$. In essence, GANs optimize the Jensen-Shannon (JS) divergence between the real and generated data distributions, and this interaction between $D$ and $G$ is characterized as a two-player zero-sum game. The resulting minimax formulation for $G$ and $D$ is defined as 

\begin{equation}
\begin{aligned}
    \min_{G}\max_{D}V(G,D) &= \mathbb{E}_{x\sim p_{\text{data}}(x)}[\log{D(x)}] \\
    &\phantom{=} + \mathbb{E}_{z\sim p_{\text{z}}(z)}[1 - \log{D(G(z))}].
\end{aligned}
\end{equation}

The joint training of $G$ and $D$ seeks to achieve an equilibrium state between the two, whereby the generator produces synthetic data of high quality that is indistinguishable from real data by the discriminator. Nonetheless, this approach is afflicted by various limitations, including mode collapse, training instability, and hyper-parameter sensitivity. In response, Arjosky et al.~\cite{mArjovsky2017} introduced a modified version of the conventional GAN loss function, which effectively reduces the unstable training dynamics, the issue of vanishing gradients, mode collapse, and convergence challenges encountered in the original GAN framework. In the proposed Wasserstein GAN (WGAN) model, the loss function is designed to optimize the Wasserstein distance, also known as Earth-Mover distance (EMD), between the real $X_r$ and generated data $X_g$ distributions, i.e.,
\begin{align}
    W(X_r, X_g) = \inf_{\gamma \sim \Gamma(X_r,X_g)} \mathbb{E}_{(x_r, x_g) \sim \gamma} [||x_r-x_g||]
\end{align}
where $\Gamma(X_r, X_g)$ is all possible joint distributions of real distribution $X_r$ and generated distribution $X_g$. Due to implementation difficulty, the Kantorovich-Rubinstein duality of EM has been proposed computing $W$ as 

\begin{equation}
\begin{aligned}
    W(X_r, X_g) &= \frac{1}{K}\sup_{||f||_{L}\le K} \mathbb{E}_{x_r \sim X_r} [f(x_r)]\\
    &\phantom{=} \quad \quad\quad - \mathbb{E}_{x_g \sim X_g} [f(x_g)]
\end{aligned}
\end{equation}
where $f$ is the set of 1-Lipschitz functions. To make the training procedure faster, Gulrajani et al.~\cite{iGulrajani2017} proposed a gradient penalty to this loss function in a method called WGAN-Gradient Penalty (WGAN-GP),

\begin{equation}
\begin{aligned}
    \min_{\theta_G}\max_{\theta_D}L(X_r,X_g) &= \mathbb{E}_{x_r\sim X_r}[D(x_r)] \\
    &\phantom{=} - \mathbb{E}_{x_g\sim X_g}[D(x_g)] \\
    &\phantom{=} - \lambda \mathbb{E}_{\hat{x}\sim \hat{X}}[(||\nabla_{\hat{x}}D(\hat{x})||_2 - 1)^2]
\end{aligned}
\end{equation}
where $\lambda$ is the penalty coefficient that controls the trade-off between original objective and the gradient penalty and $\hat{x}$ is the data points sampled from the straight line between real distribution $X_r$ and generator distribution $X_g$, specifically,
\begin{equation}
\begin{aligned}
    \hat{x} &= \alpha x_r + (1-\alpha)x_g \\
    &\phantom{=} \textrm{s.t.} \, \alpha \sim U[0,1] \\
    &x_r \sim X_r \& x_g \sim X_g.
\end{aligned}
\end{equation}

\section{Methodology}\label{sec:method}

\subsection{Sources} 
In order to identify studies that employed a non-deterministic data augmentation technique for constructing a neuronal emotional recognition model, a search was conducted across four databases: 1) ACM Guide to Computing Literature; 2) IEEE Xplore; 3) ScienceDirect, and 4) Scopus.  These databases are chosen to provide comprehensive coverage of relevant literature on computing, human-computer interaction, affective computing and brain-computing interface.

The search was conducted on November 20, 2022 with no langauge or date restrictions. Listed keywords were searched on the title, abstracts, and meta-data like tags. Table \ref{tab:keywords} lists keywords related to the research domain. The search terms were adapted to suit the specific syntax of each database. 

\begin{table}[H]
\caption{Key words search}
\label{tab:keywords}
\begin{tabular}{@{}ll@{}}
\toprule
\textbf{Domain}   & \textbf{Keywords (OR)}                 \\ \midrule
Data Augmentation and Synthesis & augment*, generat*, synthe* \\
Modality          & EEG, fNRIS                             \\
HCI               & emotion, affective                     \\ \bottomrule
\end{tabular}
\end{table}

\subsection{Inclusion and Exclusion Criteria} 
The selection criteria is set to filter out relevant literature as all retrieve papers may not be under the scope of this review. 
\begin{itemize}
    \item Exclude: Introductions, letters and comments, abstracts, books, reviews, lecture notes, PhD thesis, demonstrations
    \item Exclude: Non-English papers
    \item Include: Studies that focused on building emotion recognition models (e.g. Using Generative Adversarial Networks for Synthetic EEG data in Epilepsy classification would not be included)
\end{itemize}

The titles and abstracts of the identified articles were screened independently by two reviewers to determine whether they met the selection criteria. Any discrepancies between the two reviewers would be resolved through discussion. One hundred percent agreement was obtained.

\subsection{Limitation} 
The current search methodology suffers from a drawback in that only papers that explicitly referred to data augmentation, synthesis, or generation were considered during the identification process. Consequently, studies that may have employed these techniques to overcome data scarcity but did not explicitly state them in the abstract or title could have been overlooked in the initial search outcomes.

\section{Overview of the Selected Publications}
\label{sec:overview}
\begin{figure}
    \centering
    \includegraphics[scale=0.08]{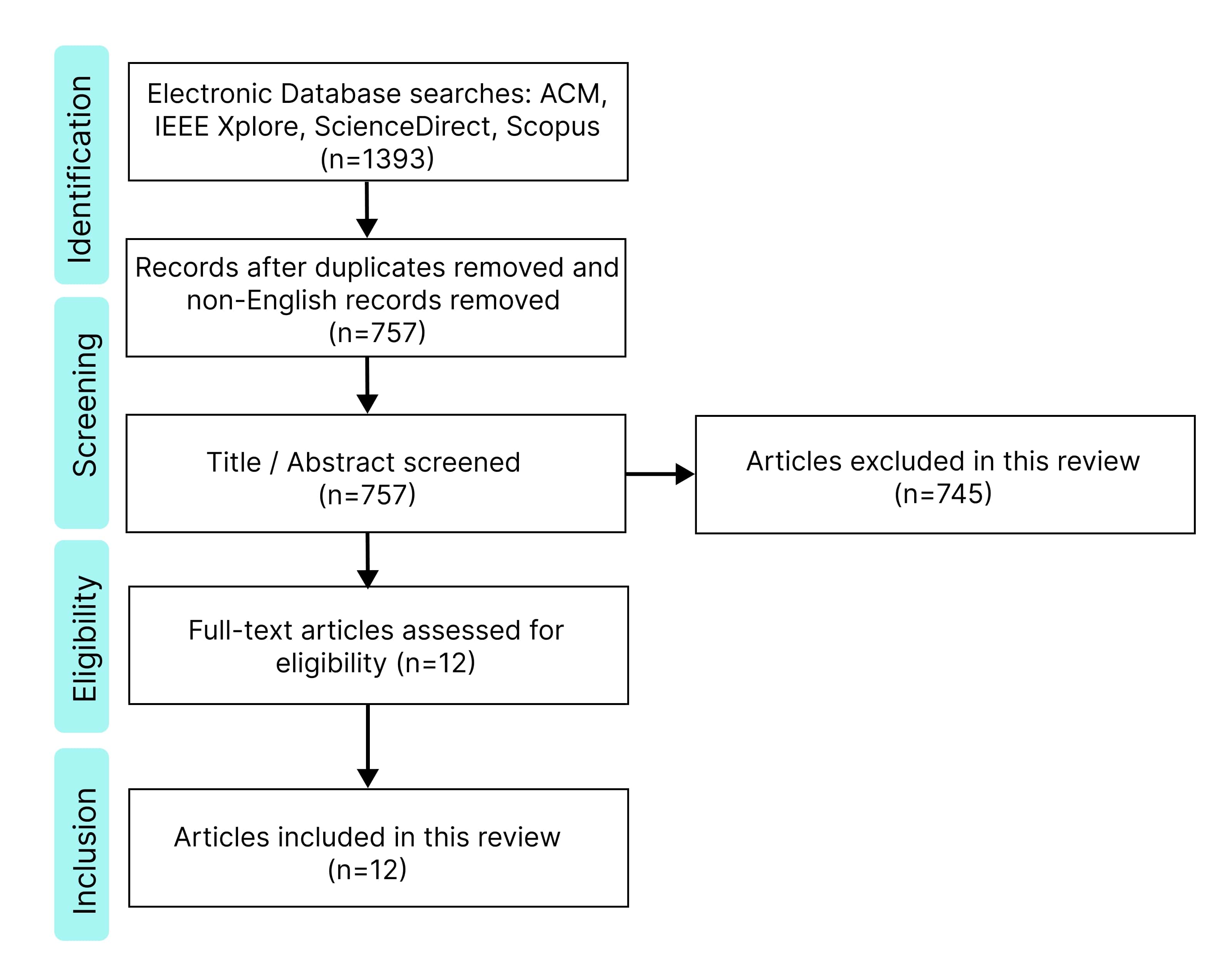}
    \caption{Flow diagram of the selection process.}
    \label{fig:flow_diagram}
\end{figure}

Figure \ref{fig:flow_diagram} illustrates the step-by-step results of the systematic review process. Our initial search returned 1393 records from electronic databases, including 181 from ACM, 162 from IEEE Xplore, 296 from ScienceDirect, and 754 from Scopus. After removing duplicates and excluding non-English papers, 757 records remained for title and abstract screening.

Based on the inclusion and exclusion criteria, 12 papers were selected for full-text review. These studies were examined closely for their relevance to the systematic review, and finally all 12 papers were included.

\section{Non-deterministic Data Augmentation Approaches}\label{sec:approaches}

This section reviews 12 selected studies that leverage non-deterministic techniques for neurophysiological data augmentation. These studies are broadly categorized based on their core methodologies into autoencoder, GAN, and hybrid approaches. Each study is evaluated regarding its data preprocessing methods, neural network designs, and classification outcomes. 

Essential terminologies are clarified for the context of this review. ``Baseline results" refers to classification outcomes prior to the inclusion of augmented data. ``Augmented" and ``synthesized" are interchangeable, denoting artificially generated data.
Three distinct data splitting strategies are identified: ``subject-dependent", ``subject-biased", and ``cross-subject". In ``subject-dependent" experiments, individual subjects have exclusive training and testing sets. ``Subject-biased" involves using a fixed number of trials from each subject for training and the remaining for testing. This could be conducted systematically  or randomly by shuffling all trials of all subjects together before separating them into train and test sets. ``Cross-subject" experiments segregate $n$ number of subjects for testing and the rest for training. In the Leave-One-Out Cross-Validation (LOOCV) experiment, as a subcategory of the cross-subject group, $n-1$ subjects are selected for training and one subject for testing. 

Lastly, classification problems addressed are either single-class or multi-class. Single-class classification (SCC), or binary classification, refers to two-label classification tasks. For example, in the DEAP dataset~\cite{sKoelstra2011}, SCC refers to a classification task in which a sample either belongs to high or low valence/arousal dimension. Alternatively, multi-class classification (MCC) refers to multi-label classification tasks. For instance, a multi-class problem for the DEAP dataset could involve categorizing samples into four distinct classes: high arousal high valence (HAHV), high arousal low valence (HALV), low arousal high valence (LAHV), and low arousal low valence (LALV). Similarly, the samples from the SEED dataset~\cite{wZheng2015} can be categorized into one of three classes: positive, neutral, or negative.




\subsection{Autoencoder}
Ari et al.~\cite{bAri2022} proposed a data augmentation technique, \textit{ELM-W-AE}, that different wavelet kernels were used as activation functions within an Extreme Learning Machine Autoencoder (ELM-AE)~\cite{lKasun2013} to enhance their emotion recognition system. The intention was to devise a rapid and straightforward method for boosting the sample size. Considering the input data as $x \in\mathbb{R}_{N\times w}$ in which $N$ is number of samples and $w$ is the number of features, and $\psi$ as wavelet kernel, the hidden layer output of the ELM-W-AE can be presented as follows:  
\begin{align}
    H_{\psi} = \begin{bmatrix} 
                \psi(a_1^Tx_1+b_1) & \dots  & \psi(a_K^Tx_1+b_K)\\
                \vdots & \ddots & \vdots\\
                \psi(a_1^Tx_N+b_1) & \dots  & \psi(a_K^Tx_N+b_K)
             \end{bmatrix}_{NxK}
\end{align}
in which $K$ is the number of the nodes in the hidden layer, $(a_k, b_k)$ is the fixed input weights and biases of hidden neurons. Similar to the ELM algorithm, the output weights $\beta$ of the ELM-W-AE can be calculated with $\beta=H^{'}X$ in which $H^{'}$ is the Moore-Penrose inverse of $H$~\cite{gHuang2004}. Finally, the output can be derived as $\hat{X}=\beta H$. 

This study utilized the GEMEEMO Dataset~\cite{tAlakus2020-2}, which comprises EEG data collected from 28 participants using a 14-channel EEG device. The EEG signals were transformed into 2D scalogram images through the continuous wavelet transform (CWT) with a scale parameter of 12 and analytic Morse wavelet. 
The output was normalized and resized to a dimension of $224\times224$ for input into the model. Six different wavelet functions (Gaussian, Morlet, Mexican, Shannon, Meyer, and GGW) were employed within the ELM-W-AE to each expand the training set by a factor of six. The results from the different wavelet functions were then compared.
ResNet18 model~\cite{kHe2016} was used to detect four emotional states in a MCC experiment via a subject-biased approach in which subjects and trials were merged. Notably, the proposed method showed an accuracy improvement of approximately 20\% (with accuracy equaling 77.66\%) over the baseline, with the GGW activation function yielding the highest performance scores (accuracy = 99.57\%).

Drawing inspiration from the successful language model, \textit{word2vec}~\cite{tMikolov2013}, Bethge et al.\cite{dBethge2022} introduced framework named \textit{EEG2Vec}. The proposed model was designed to provide generative-discriminative representation for EEG data through a variational feature encoder that generalizes across different subjects ($p$) and emotion categories ($y$).  It incorporates an adjacent decoder to generate new data samples from a variational probability distribution and uses latent features to signify corresponding emotional states. This latent construct has a dual function: generating samples and predicting affective states. 
A novel loss function, defined below, is introduced for the simultaneous optimization of the conditional $\beta$-VAE and the classifier. 
\begin{align}
    \mathcal{L}_{EEG2Vec} = \mathcal{L}_{\beta-cVAE} + \lambda \underbrace{\mathbb{E}[-\log r_{\varphi}(y|z)]}_{\mathcal{L}_{classifier}}
\end{align}
The $\mathcal{L}_{classifier}$ term represents the cross-entropy loss of the emotion recognition system. The classifier, parameterized by $\varphi$, uses the estimated latent feature ($z$) from the $\beta$-cVAE to predict the emotional label ($y$). The parameter $\lambda>0$ controls the weighting between the generative and discriminative components of the model. The $\mathcal{L}_{\beta-cVAE}$ term is defined as follows:
\begin{align}
    \mathcal{L}_{\beta-cVAE}(\theta, \phi) = \begin{aligned}[t] &-\mathbb{E}_{z \sim q_{\theta}(z|x_{i})}[\log_{p_{\phi}}(x_{i}|z,y,p)] \\
    &+ \beta \mathbb{KL}(q_{\theta}(z|x_{i}) || p(z) )
    \end{aligned}
\end{align}
The decoder is conditioned on the emotional label ($y$) and the specific subject ($p$), while the encoder aims to learn latent representations ($z$) that are independent of these factors. The latter are provided as input to the decoder for input reconstruction ($X$). A parameter $\beta > 1$ steers the cVAE towards a standard normal distribution.
The authors assessed the performance of their framework using the SEED dataset. 
A 155-segment of each trial was chosen and the data was divided into non-overlapping windows of 2 seconds each. A holdout testing set of 10\% was initially separated, while the remaining 90\% was split into training and validation sets using 5-fold cross-validation. Equal representation of affective states (classes) and subjects was maintained across the training, validation, and testing sets. By implementing the EEG2Vec model to generate synthetic data, the authors observed that combining the original data with 20\% synthetic data increased MCC accuracy from 66\% to 69\%. While the fully-discriminative EEGNet baseline model achieved an affective state prediction accuracy of 77.27\% on the testing set, authors believed that the proposed architecture has the capability to learn both efficient (lower dimensionality with $z$) and expressive (retaining useful properties with $z$) representations.

Luo et al.\cite{yLuo2020} sought to examine the impact of the quality of generated data on the performance of an emotion recognition system. The authors used a conditional VAE and conditional WGAN (cWGAN), introducing emotional labels $y$ into the data generation process. Two strategies were compared: the first entailed adding all generated data (full usage), and the second involved adding only the highest quality generated data (partial usage), termed as selective VAE (sVAE) and selective CWGAN (scWGAN). To evaluate data quality, a classifier (Support Vector Machine (SVM) or Deep Neural Network (DNN) with shortcut layers), trained on the original dataset, was used to classify the generated data. Data with a classification confidence over 0.4 were added to the training set, and this process was repeated until enough generated data was accumulated. 
The network architectures of both the cWGAN and DNN were optimized using grid search for relevant hyperparameters. 
MCC experiments were conducted on the SEED and DEAP datasets. 
DE features were extracted from five frequency bands ($\delta$, $\theta$, $\alpha$, $\beta$, and $\gamma$) in the SEED dataset, and from four frequency bands (excluding $\delta$) in the DEAP dataset. Different numbers of generated DE data were experimented with, ranging from 0 to 20,000.
In the VAE application, the largest accuracy improvement (4.2\% and 4.4\%) from baseline for SEED and DEAP dataset were achieved using sVAE with DNN as the classifier. The authors observed that classification accuracy declined when an excessive amount of generated data was appended for both datasets. Hence, they recommended limiting the number of generated data points to less than ten times the size of the original dataset to achieve optimal model performance. 
Simiarly, in the GAN application, partial usage of data exhibited superior performance in both datasets, achieving an accuracy improvement of 3.4\% with SVM and 1.9\% with DNN on the SEED Dataset. However, the authors noted that the full usage strategy enabled faster convergence of the cWGAN model. 

\subsection{Generative Adversarial Networks (GANs)}
An increasing number of researchers have turned to GANs for generating synthetic neuronal signals, motivated by their impressive capacity to output high-quality data compared to other generative models such as AEs. Specifically in the realm of affective computing, the complexity and non-stationarity inherent in neurophysiological data necessitate sophisticated models like GANs. These models may be more adept at capturing the intricate latent representations of this data type, enabling the synthesis of data that accurately reflects its nuanced characteristics.

Addressing the imbalance samples between different emotional categories, Pan et al.~\cite{bPan2021} proposed an EEG-based emotion recognition system that leverages a vanilla GAN to increase the training samples so that the samples in each category was the same as the largest number in different categories. Relying on PSD features of EEG signal, they proposed \textit{PSD-GAN}, which generates PSD features from EEG signals. The proposed model included a generator $G$ network featuring three linear layers and three activation functions (two ReLU and one Tanh) and a discriminator $D$ network with two linear layers and two activation functions (a LeakyReLU and a Sigmoid function).
The authors validated the performance of their method using the DEAP dataset. They initially mapped the 32 EEG channels onto a 2D sparse map based on electrode positions in the 10-20 system, removed the baseline of each trial, and segmented the data into 2-second windows with a 1-second overlap. Each segment retained the original trial label. The PSD feature was then extracted from four frequency bands ($\theta$, $\alpha$, $\beta$, and $\gamma$).
The evaluation was conducted with only 12 subjects. Both SCC and MCC were performed using two different classifiers on the valence and arousal dimensions. The authors utilized a 10-fold cross-validation methodology for each subject (subject-dependent) and all subjects and trials together (subject-biased). Two convolutional neural networks (CNNs) were employed to classify the extracted PSD features. We report the average results of both classifiers. 
In the subject-dependent evaluation, the augmentation technique increased accuracy by 5.25\% and 6.83\% along the valence and arousal dimensions, respectively, for SCC. In MCC, an average of 10.92\% increase was achieved. For the subject-biased evaluation, the proposed pipeline yielded improvements of 6.5\% and 6.71\% along the valence and arousal dimensions, respectively, for SCC, and 14.47\% for MCC. A similar trend in classification accuracy improvement was also observed when deploying the model on MAHNOB-HCI database~\cite{jLichtenauer2011}. Despite these improvements, the authors noted challenges in training the GAN network to generate PSD features, with the quality of the generated samples being unstable.

Considering the impact of data augmentation on emotion recognition systems, Bhat et al.~\cite{sBhat2021} employed WGAN-GP to examine the usability of the GANs in generating augmented datasets to improve the classification performance for emotion recognition tasks. They introduced a framework that involves training a WGAN-GP for 100 epochs, saving the generated samples after each epoch. High-quality data, defined as those generated with WGAN-GP losses (for both generator and discriminator) close to zero, were added to the original dataset to evaluate the classification model. Specifics regarding the structure of the utilized WGAN-GP were not provided by the authors.
The evaluation was conducted on the DEAP dataset whereby the signals were divided ten 6-second segments using a sliding window. Nine features were extracted, including mean, median, maximum, minimum, standard deviation, variance, range, skewness, and kurtosis, from both the entire trial and each segment. 
A cross-subject approach was taken with 22 subjects for the training set, and 10 remaining subjects in the testing set. 
WGAN-GP was only implemented on the training set, with $G$ fed by randomly sampled noise with the length of $256$ and output shape of $(40,99)$. 
Due to the limited amount of data, data generated at epochs 89, 91, 97, 100 were chosen for augmenting the data by a factor of 2, 3, and 4. Three different classifiers (KNN, SVM, and CNN) were tested in the SCC experiment. It was observed that KNN with augmentation outperformed other models by having the highest accuracy of 62.5\% (5.0\% increase from baseline), 59.25\%(2.5\% increase), 61.5\%(2.5\% increase) and 71.5\% (7.5\% increase) along valence, arousal, dominance and liking dimensions. 

Building on the original WGAN-GP and to generate multiple emotion categories, Luo et al.~\cite{yLuo2018} introduced data labels as auxiliary conditional information to WGAN-GP in order to improve the performance of the emotion recognition system. 
To achieve this, data labels ($Y_r$) were fed into both the discriminator $D$ and generator $G$ in WGAN-GP, and the loss function was as follows:
\begin{equation}
\begin{aligned}
    \min_{\theta_G}\max_{\theta_D}\mathcal{L}(X_r,X_g, Y_r) &= \mathbb{E}_{x_r\sim X_r, y_r \sim Y_r}[D(x_r|y_r)] \\
    &\phantom{=} - \mathbb{E}_{x_g\sim X_g , y_r \sim Y_r}[D(x_g|y_r)] \\
    &\phantom{=} - \lambda \mathbb{E}_{\hat{x}\sim \hat{X}, y_r \sim Y_r}[(||\nabla_{\hat{x}|y_r}D(\hat{x}|y_r)||_2 - 1)^2]
\end{aligned}
\end{equation}
The proposed framework was evaluated on two public datasets, DEAP and SEED. To find the configuration of the proposed cWGAN-GP, a grid search was performed on the number of layers between 3 to 5 for both the discriminator and generator, while the number of nodes for each hidden layer was fixed as 512 and 256 for SEED and DEAP dataset, respectively.  In both datasets, each EEG signal was segmented into 1-second non-overlap chunks, and each segment was decomposed to its frequency bands for which the DE feature was extracted. 
To build training test, 24 (out of 40) trials for the DEAP dataset and 9 (out of 15) were randomly selected from each subject; and the remaining trials were used as testing set. 
Only high-quality generated data was added to the training set. 
The quality of the generated data was assessed using three metrics: (i) Discriminator loss -- ideally $\mathcal{L}_D \approx0$; (ii) Maximum Mean Discrepancy (MMD) between the distribution of generated $X_g$ and real data $X_r$ -- ideally $D_{MMD}\approx0$; (iii) visual inspection of 2D mapping of $X_g$ using t-Distributed Stochastic Neighbor Embedding (t-SNE). 
The authors used an SVM classifier to evaluate the impact of their proposed augmentation framework and the size of the augmentation on their emotion recognition system. Over a subjective MCC on SEED dataset, it was observed that classification accuracy improved by 2.97\% on average over all categories. Similarly, the DEAP dataset provided 9.15\% and 20.13\% improvements along arousal and valence dimensions, respectively. In both experiments, the best improvements occurred for augmentation size equals to $1\times\textit{Dataset}$. 


In another study, to investigate the similarities and differences of EEG and eye movement signals between Chinese and French on emotion recognition tasks, Gan et al.~\cite{lGan2019} utilized the cWGAN to address the data scarcity in their study. The task was to classify positive, neutral, and negative emotions, and they performed this analysis on a dataset composed of six randomly selected Chinese participants from the SEED database and six additional French participants from their own experiments. For the purpose of this review, the impact of augmentation on EEG data was analysed. The authors followed a similar approach as~\cite{yLuo2018} to generate new EEG samples. 
To analyze the EEG data, a short-time-frequency transformation (STFT) with a 4-second non-overlapping window was applied to the data to transform the data into time-frequency space. DE features were extracted from transformed data as input for the cWGAN-GP network. Both generator and discriminator were fixed with four hidden layers, each with 512 nodes optimised through grid search. 
The authors used \textit{Adam} optimizer ($lr=1e-3$) and penalty value of $\lambda=10$. A linear SVM was used to evaluate the MCC classification performance over subject-biased 5-fold cross-validation. During the cross-validation test, generated data was combined with each of the 5-fold training datasets. By incorporating triple-generated data ($3\times \text{Dataset}$) into the original training set, the classification accuracy saw an improvement of 2.09\%, increasing from 49.97\% to 52.06\%.

In a separate study but with a similar framework as~\cite{yLuo2018}, Kalashami et al. \cite{mKalashami2022} applied cWGAN-GP to enhance emotion recognition accuracy within the SCC experiment. They chose the DEAP dataset to perform their affective framework. Following ~\cite{zYin2017}, researchers extracted the average PSD for each channel individually for five frequency bands, including $\theta [4\sim8]$, slow-$\alpha [8\sim10]$, $\alpha [10\sim12]$, $\beta [12\sim30]$ and $\gamma [30\sim45]$. Additionally, they extracted the difference of the paired channel from the right and left hemispheres in four frequency bands. Zero-crossing rate, mean, and variance in the total frequency range were also computed. The authors followed a subject-biased evaluation method and only considered using only high-quality generated samples. 
The generated data were evaluated via two methods before appending to the original dataset: (i) comparison of real data and generated data distribution, and (ii) analysis of the loss functions of generator $G$ and discriminator $D$. High-quality generation corresponded to low $G$ and high $D$ losses. Training was deemed complete upon loss value stabilization. Furthermore, Principal Component Analysis (PCA) was utilized for dimensionality reduction and for isolating two most significant eigenvalues to visually evaluate the generated data. Training ceased when the dispersion of the real and generated data aligned.
The model achieved an accuracy of 68.2\% with a 3.9\% increase from the baseline for the arousal dimension and 64.3\% with a 4.2\% increase for valence, using the SVM classifier. Furthermore, employing a costumed DNN classifier consisting of two hidden layers (256,128) with a dropout layer and ReLU activation, authors observed an accuracy of 71.9\% (6.5\% improvement) for arousal and an accuracy of 67.4\% (3.1\% increase) for valence. 

Building on Luo et al.'s work~\cite{yLuo2018}, Zhang et al.~\cite{aZhang2021} proposed multi-generator conditional WGAN (MG-cWGAN) in order to mitigate the mode collapse and increase the variation in generated samples. This approach employs a mixture generator that approximates input feature distribution by generating a corresponding mixture distribution. This allows the generator to learn the feature patterns of various distributions~\cite{qHoang2018}. The authors incorporated label-based conditional constraints to direct the feature generation process to avoid the high computational cost and divergence that can arise from using multiple generators. The discriminator learns real data based guided by the label constraints and further forces the generators to learn specified features. The generator, composed of input layers and parameter-sharing layers, processes a joint input of a prior distribution with guidance labels. Similarly, the discriminator accepts a joint input of real data with labels and artificial data with guidance labels. The discriminator returns gradients to guide the generator to learn features of the real data. This results in the following modification of the cWGAN-GP generator and discriminator loss function:
\begin{equation}
\begin{aligned}
    \max_{\theta_D}\mathcal{L}(X_r,X_g, Y_r) &= \mathbb{E}_{x_r\sim X_r, y_r \sim Y_r}[D(x_r y_r)] \\
    &\phantom{=} - \mathbb{E}_{x_g\sim X_g , y_r \sim Y_r}[D(x_g y_r)] \\
    &\phantom{=} - \lambda \mathbb{E}_{\hat{x}\sim \hat{X}, y_r \sim Y_r}[(||\nabla_{\hat{x}y_r}D(\hat{x} y_r)||_2 - 0)^2]\\
    \min_{\theta_{G_{1:N}}}\mathcal{L}(X_g, Y_r) = \sum_{i}^{N} -\mathbb{E}_{x_g\sim X_g, y_r\sim Y_r}[D(x_{g_i}| y_r)] 
\end{aligned}
\end{equation}
where $\theta_{G_{1:N}}$ are the number of generators and input noise vector $Z$ to each generator $G_i$ sampled from $P_z$ is connected to the label information as $(z|y_r)$ to generate features belong to $y$ label. Modification of the gradient penalty term to a zero-centered gradient penalty term enhances the convergence of the model. In addition, to increase the training efficiency of the model, a method~\cite{sWang2021} was adopted for sharing all the parameters of the generator $G_{1:N}$ except those of the input layer. The authors conducted comparative subjective experiments using cGWAN-GP and MG-cWGAN on the SEED dataset. All trials of a subject were divided into training and test sets, and the performance of the augmentation framework was evaluated in the 5-fold cross-validation setup. DE features of each trial were used as input features for classification and augmentation. 
To optimize the configuration for generator $G$ and discriminator $D$, a grid search was conducted on the learning rate, the number of network layers, and the batch size of the deep neural network classifier. The influence of the number of hidden layers on the MG-cWGAN model was found to be minimal, whereas the cWGAN model exhibited more pronounced sensitivity. Specifically, cWGAN, which consisted of only three hidden layers, displayed instability during the later stages of training. In an experiment on MG-cWGAN, it was observed that a batch size of 128 resulted in the slowest training speed and lowest stability. A batch size of 256 yielded satisfactory convergence with low divergence, while a batch size of 512 led to fast and satisfactory convergence. The approach of adding generators for various types of signals marginally enhanced network stability but significantly reduced the model's convergence speed. Consequently, the utilization of zero-centered gradient penalties in MG-cWGAN proved instrumental in substantially improving the network's convergence performance. When comparing the discriminator loss (Wasserstein distance) of cWGAN-GP and MG-cWGAN under identical parameters, it became evident that the MG-cWGAN model achieved notable improvements in convergence speed and stability, surpassing those of the cWGAN-GP model. 
The quality of the generated data was assessed using MMD, the Wasserstein distance in discriminator loss, t-SNE, and a semi-supervised classifier. Through visualization using t-SNE, authors observed that the data generated by MG-cWGAN was close to the corresponding real data; hence, carry more information about the real data than that from cWGAN-GP. The authors employed a semi-supervised self-training framework for evaluating the augmentation technique in which the labeled initial data set ($X_{\text{train}}, Y_{\text{train}}$) with the "pseudo-labeled" generated dataset ($X_{\text{g}}, Y_{\text{c}}$) is used for training a classifier $C_{\text{int}}$ which later is used to classify the unlabeled data and produced the "pseudo-labeled" dataset ($X_{\text{g}}, Y_{\text{c}}$). These two datasets together were used to train the final classifier $C$, whose performance was evaluated on the initial test set. This performance is also used as an indicator of the quality of augmented data. They used KNN and SVM as classifiers in this process. Using SVM, the classification performance for that of MG-cWGAN has improved by about 1\% from baseline, reaching 84\% accuracy; however, a decrease in performance using the cWGAN-GP data augmentation method was observed. Using KNN, while both cWGAN-GP and MG-cWGAN frameworks improved the classification accuracy in comparison to the baseline, cWGAN-GP increased the KNN accuracy by 3\%, reaching 80\%, and MG-cWGAN increased it by 1.5\% reaching to 78.5\%.   

Zhang et al.~\cite{Zhang2022_GANSER} took on a different approach in using GAN for augmenting the data. The authors introduced Adversarial Augmentation Network (ANN) called Generative Adversarial Network-based Self-supervised Emotion Recognition (GANSER). 
The proposed process begins by transforming 1D EEG signals into a 2D $9\times9$ sparse map using the 10-20 system. This results in a new representation of the EEG channels, from $e\in\mathbb{R}^{128\times32}$ to $e\in\mathbb{R}^{128\times9\times9}$, where $e$ represents a sample of the dataset. The Masking Transformation operation, achieved through the random sampling of a uniform distribution matrix of the same size as the EEG signal $ e\in\mathbb{R}^{128\times9\times9}$, to partially cuts out signals. The amount of signal removed depends on the augmentation factor ($\tau$) sampled from a uniform distribution ($U\sim[\tau_{min}, \tau_{max}]$), allowing the generation of a variety of augmented samples.
\begin{equation}
  \delta(e_{ijk}, \tau) =
    \begin{cases}
      e_{ijk} & \quad r_{ijk}>\tau \\
      0 & \quad r_{ijk}<\tau\\
    \end{cases}       
\end{equation}
where $\delta(e,\tau)$ is the simulated EEG signals transformed from the original signal $e$ based on the augmentation factor. The larger $\tau$ factor removes a larger part of the original EEG signal. 
A GAN is then introduced. Contrary to the Masking Transformation, which operates at the signal value level, the GAN focuses on learning the feature-level distribution of realistic EEG signals. The aim is for the generated samples to retain the natural features of real data while also introducing diversity. 

To mitigate the stability challenges typically encountered with traditional GANs, the authors implemented a variation of WGAN-GP, integrating it with the random masking augmentation approach as outlined below:
\begin{equation}
\begin{aligned}
    \min_{D}\mathcal{L}_{D} &= \mathbb{E}_{e\sim P_{e}}[D(G(\delta(e, \tau)))] - \mathbb{E}_{e\sim P_{e}}[D(\delta(e,\tau))] \\
    &\phantom{=} + \lambda_{p} \mathbb{E}_{\hat{e}\sim P_{\hat{e}}}[(||\nabla D(\hat{e})||_2 - 1)^2]\\
    \min_{G}\mathcal{L}_{G} &= -\mathbb{E}_{e\sim P_{e}}[D(G(\delta(e,\tau))] 
\end{aligned}
\end{equation}
where $P_{e}$ is the distribution of the original EEG signal. 
The generator utilizes a variant of UNet consisting of an encoder-decoder structure with skip connections. The encoder, comprising four 2D-convolutional layers with LeakyRelu activation function, downsamples the input signal and maps its spatiotemporal information into a latent representation. Then, the decoder, consisting of three 2D-deconvolutional layers, upsamples the latent representation to high spatiotemporal resolution, synthesizing the missing values to generate new EEG samples. Skip connections are designed between the convolutional and symmetric de-convolutional layers to merge shallow feature maps, thereby helping de-convolutional layers supplement high-resolution details.
The discriminator, a novel STNet, is proposed to analyze the complex spatiotemporal features of EEG signals. This network consists of three 2D-convolutional layers, a separable convolutional layer, and an Inception block. After high-level spatiotemporal feature maps are extracted by the convolutional layers, a separable convolutional layer is employed to decouple the modeling of spatiotemporal information and capture spatial and temporal correlations of the extracted feature maps, respectively. The final block of the proposed network, the Inception block, is used to extract multi-scale feature maps. By fusing these feature maps, the pattern of emotions related to both multiple electrode signals and local electrode signals can be adaptively captured. 
The authors propose a multi-factor training network for emotion classification, inspired by a self-supervised learning framework. Given the influence of the augmentation factor $\tau$ on the feature distribution of generated samples, surrogate confidence values are used to limit the distribution disparity between real and augmented samples. For large $\tau$ values, the generator $G$ may fail to retain the original EEG signal's features,  necessitating a constraint on the augmented signals' feature distribution.
Therefore, a loss function, as defined below, is introduced, using cross-entropy loss to assign weights that manage the distribution between real and augmented samples based on the respective surrogate confidence.
\begin{equation}
\begin{aligned}
    \mathcal{L}_C(\tau) &= -\frac{1}{n}\sum_{i=0}^{n}y_{i}\log(C(e_i)) \\
    &\phantom{=} + \frac{\lambda_a}{n} \sum_{i=0}^{n}(1-\tau_i)||C_x(G(\delta(e_i, \tau_i)))-C_x(e_i)||_{2}^2\\
\end{aligned}
\end{equation}
$C$ represents the classifier, $C_x$ is the classifier without the last fully connected layer, $e_i$ and $y_i$ are sample EEG signal and its label, respectively, $n$ is the number of samples in a mini-batch, $\lambda_a$ is a hyper-parameter for regularization. The initial part of the loss function is the cross-entropy loss, and the second part includes the impact of the augmentation factor $\tau_i$ using surrogate confidence $(1-\tau_i)$. The feature distribution difference between the original EEG signals and the corresponding augmented signals is denoted by $||C_x(G(\delta(e_i, \tau_i)))-C_x(e_i)||_{2}^2$.

In this pipeline, the separated stages of augmentation and training are joined together as an end-to-end pipeline. 
A batch of samples is first augmented using the generator $G$ in AAN, and then these samples are used to optimize the classifier $C$ with the proposed loss function $L_C$ in the current batch. This approach is more efficient as it avoids the cost of saving and reloading. Moreover, augmented samples are regenerated between epochs in real-time. Benefiting from the randomness of the Masking Transformation operation, augmented samples of the corresponding batch between epochs are different, but both sampled near the real distributions. This allows randomly synthesized EEG signals to approximate the distribution of real EEG signals without a number limitation, thus avoiding overfitting on a preset number of augmented samples.
The authors used 5-fold cross-validation subjectively to evaluate their proposed pipeline on three public datasets. 
Using the DEAP dataset, the EEG signals were first segmented into 1-second segments using a non-overlapping sliding window and baseline removal was applied. The resulting number of samples totaled to $40\times60=2400$. The proposed method shows great SCC performance of over 93.52\% and 94.21\% along valence and arousal dimensions, respectively. In addition, for subjective MCC, the proposed GANSER can correctly classify 89.74\% EEG signals. In the case where no data augmentation is utilized, the proposed classifier can achieve an accuracy of 91.39\% and 92.22\% at valence and arousal dimensions. With subject-independent evaluation on the DEAP dataset, authors reported 49.36\% and 55.12\% accuracy over valence and arousal dimensions, respectively. 
Similarly for the SEED dataset, the signals were first segmented and baseline was removed. MCC accuracy was 97.7\%. And on the DREAMER dataset, the proposed method achieved 85.28\% and 84.16\% accuracy over valence and arousal dimensions. 

To further evaluate the synthetic data generated, authors adapted Frechet Inception Distance (FID) and designed Frechet STNet Distance (FSTD). They first trained the proposed GAN  on 80\% training samples to distinguish different emotions. EEG samples are then fed into the learned STNet to extract the output of the penultimate fully connected layer. Authors then computed the mean $\mu$ and covariance $\Sigma$ of real and generated samples and then measured the Wasserstein distance between two multivariate normal distributions, $N(\mu, \Sigma)$ and $N(\mu_g, \Sigma_g)$ as follows:
\begin{equation}
\begin{aligned}
    \text{FSTD} &= ||\mu - \mu_g ||^2 + Tr(\Sigma - \Sigma_g - 2\sqrt{\Sigma\Sigma_g})
\end{aligned}
\end{equation}
where a small value of FSTD indicates a high similarity between the generated samples and real data distribution.

\subsection{Hybrid Generative Models}
Motivated by the problem of data scarcity and drawing upon the advantages of the VAE and GAN networks, Bao et al.~\cite{gBao2021} proposed a model named VAE-D2GAN. This model leverages the VAE network to map real data into a latent distribution, which is subsequently fed into a generator for more accurate and efficient learning of the actual data distribution. The GAN component, inspired by D2GAN\cite{tNguyen2017}, comprises a generator, $G$, and two discriminators, $D_1$ and $D_2$. The generator $G$ takes random variable $z_p$ as input, where $z_p\sim N(0,1)$. 
The relationship between $G$, $D_1$ and $D_2$ can be modelled as follows:
\begin{equation}
\begin{aligned}
    \min_{G}\max_{D_1, D_2}\mathcal{L} &=\alpha \mathbb{E}[logD_1(x_{real})]+\mathbb{E}[-D_1(G(z_p))] \\
    &\phantom{=} +\mathbb{E}[-D_2(x_{real})]+\beta \mathbb{E}[logD_2(G(z_p))]
\end{aligned}
\end{equation}
in which $\alpha$ and $\beta$ were used to make the training more stable and  $0<\alpha,\beta \leq 1$). Additionally, the discriminators have the same network structures, but their parameters are not shared. In the D2GAN component, the loss function of the generator and two discriminators of the proposed framework is as follows:
\begin{equation}
\begin{aligned}
    \mathcal{L}_G &=-\mathbb{E}[D_1(G(z))]-\mathbb{E}[D_1(G(z_p))]\\
    &\phantom{=} + \beta\left(\mathbb{E}[\log D_2(G(z))]+\mathbb{E}[\log D_2(G(z_p))]\right) \\
    \mathcal{L}_{D_1} &= \alpha \mathbb{E} [\log D_1(x_{real})] + 
    \mathbb{E} [-D_1(G(z))] + \mathbb{E} [-D_1(G(z_p))] \\
   \mathcal{L}_{D_2} &= \beta \left(\mathbb{E} [\log D_2(G(z_p)))] + \mathbb{E} [\log D_2(G(z))]\right) \\
    &\phantom{=} + \mathbb{E} [-D_2(x_{real})] \\
\end{aligned}
\end{equation}a
in which $z$ is the latent distribution of VAE and $G(z)=p(x_{real}|z)$. As a result of the game between the generator and the two discriminators in D2GAN, the distribution of generated and real samples will be close to each other, preventing the mode collapse and ensuring the diversity in the generated samples. Finally, adding up VAE loss function $\mathcal{L}_{VAE}$ to the D2GAN loss functions will build the loss function of the proposed framework:
\begin{equation}
\begin{aligned}
    \min_{E,G}\max_{D_1, D_2}\mathcal{L}(E, G, D_1, D2) = \mathcal{L}_{KL} + \mathcal{L}_{rec} + \mathcal{L}_G + \mathcal{L}_{D_1} + \mathcal{L}_{D_2}
\end{aligned}
\end{equation}
In which $\mathcal{L}_{KL}$ and $\mathcal{L}_{rec}$ are the KL divergence and loss functions of the VAE component of the framework. Given the final loss function, VAE-D2GAN becomes a four-player game optimized by the encoder, generator, and two discriminators. 
The authors implemented the pipeline on SEED and SEED-IV databases. EEG data from SEED database was divided into 1-second non-overlapping segments, providing each subject with 3,394 samples. Data are classified into three emotional categories. The data from SEED-IV were divided into 4-second segments without overlapping, giving 851 samples per subject, and were classified into four categories (happy, neutral, sad, and fear). In both datasets, the DE feature was extracted for five frequency bands from each segment. Consistent with prior research~\cite{gBao2021-2}, channel spatial information was converted into a 2D sparse map via polar coordinate projection. For locations devoid of DE features (not attributable to channels), estimates were made based on the existing DE feature, employing the Clough-Tocher scheme interpolation.  
This yielded a topology-preserving DE (TP-DE) map of dimension $32\times32\times5$ for each data point. MCC experiments via a subject-dependent approach. 
The authors used a Deep Neural Network (DNN) introduced in ~\cite{gBao2021-2} as the classifier, and its performance was compared with SVM, VGG16, AlexNet, and ResNet18. For data augmentation, the authors proposed a strategy wherein $N$ distinct augmentation models with the same structure but different trained parameters were created, one for each emotional category. Finally, all the generated and real samples would be pooled together to train the classifier.
In comparative evaluation with models like VAE, WGAN, DCGAN, VAE-GAN, and D2GAN, the proposed augmentation approach demonstrated superior performance across both datasets, with accuracies of 92.5\% and 82.3\% for SEED and SEED-IV, respectively. This represents a 1.5\% and 3.5\% increase over the baseline results, with the inclusion of 2,000 and 10,000 generated samples, respectively. While different classifiers showed varying performances, SVM appeared less influenced by data augmentation. Conversely, the impact of augmented data was more pronounced in deep networks, particularly for smaller datasets.
To validate the effectiveness of the proposed method against other models, the authors employed a \textit{t}-test. They found that synthesized samples produced by D2GAN, VAE-GAN, and VAE-D2GAN were statistically indistinguishable from real samples ($p > 0.05$), implying that these models effectively learned the actual data distribution. Notably, synthetic data generated by VAE-D2GAN had the highest correlation with the real data ($p = 0.9334$).
The authors also used three metrics — IS, FIS, and MMD — to evaluate the quality of the generated samples. WGAN was found to have a higher propensity for mode collapse, as indicated by its elevated IS and FID values. Higher MMD and FID values for VAE suggested the poorer quality of the generated samples. Conversely, VAE-D2GAN, with the lowest MMD and FID values, demonstrated superior quality and diversity compared to the other approaches evaluated.

Expanding on their previous research on cWGAN~\cite{yLuo2018}, Luo et al.~\cite{yLuo2019} proposed the Conditional Boundary Equilibrium GAN (cBEGAN) aimed at synthesizing DE features from noise. A salient attribute of cBEGAN is its robustness to the typical instability and slow convergence issues seen in conventional GANs. The discriminator within cBEGAN adopts an AE structure -- an encoder, which extracts latent features from the input data, and a decoder, which reconstructs the input from these latent features. The discriminator's primary function is to equate the distribution of reconstruction losses between real and generated data using the Wasserstein distance metric. The reconstruction loss is defined as the $L_1$ or $L_2$ distance between input and reconstructed data $L_r(x)=|x - D(x)|^\upeta$ where $x$ is either real data or generated data distribution, $D$ is the discriminator (AE) and $\upeta \in 1,2$. This leads to the following loss function:
\begin{equation}
\begin{aligned}
    \min_{\theta_G}\max_{\theta_D}\mathcal{L}(X_r,X_g) &= -\mathbb{E}_{x_r\sim X_r}[L_r(x_r)] + \mathbb{E}_{z\sim Z}[L_r(G(z))] \\
    &\phantom{=} -\mathbb{E}_{x_r\sim X_r}[L_r(x_r)] + \mathbb{E}_{x_g \sim X_g}[L_r(x_g)]
\end{aligned}
\end{equation}
In order to maintain the balance between the generator and discriminator losses, a hyper-parameter $\gamma \in [0, 1]$ was defined as
\begin{equation}
\begin{aligned}
    \gamma = \frac{\mathbb{E}[L_r(G(z))]}{\mathbb{E}[L_r(x_r)]}
\end{aligned}
\end{equation}
and assigned $\gamma=0.75$. Similar to cWGAN, by adding label distribution $Y_r$, cBEGAN loss function can be written as follows:
\begin{equation}
\begin{aligned}
    \min_{\theta_G}\max_{\theta_D}\mathcal{L}(X_r, X_g, Y_r) &= \mathbb{E}_{x_g\sim X_g, y_r \sim Y_r}[L_r(x_g|y_r)] \\
    &\phantom{=} - \mathbb{E}_{x_r\sim X_r , y_r \sim Y_r}[L_r(x_r|y_r)] \\
    &\phantom{=} + \kappa_t \mathbb{E}_{x_g\sim X_g, y_r \sim Y_r}\\
    \kappa_{t+1} = \kappa_{t} + \lambda(\gamma L_r(x_r) - L_r(G(z)))
\end{aligned}
\end{equation}
where $\kappa \in [0,1]$ controls the proportion of $L_r(G(z))$ during the gradient descent. Initial values for these parameters were $\kappa_0 = 0$, $\gamma_{k}=0.001$. In addition, authors defined the convergence indicators as follows:
\begin{align}
    M_{global} = L_r(x_r) + | \gamma L_r(x_r) - L_r(G(z))|
\end{align}
The augmentation approach proposed was evaluated on two public datasets: SEED and SEED-V, utilizing both cWGAN~\cite{yLuo2018} and cBEGAN. The feature extraction procedure comprised the extraction of DE features (from 4-second windows via STFT) across five frequency bands. They conducted k-fold cross-validation for each subject based on the number of emotional labels present in the dataset ($k=5$ and $3$ for SEED and SEED-V respectively). To optimize the network structure of generative models, a grid search was conducted on the number of network layers (ranging between 2 and 4) and hidden nodes (ranging between 50 and 600). For both the generator and discriminator, the number of layers varied between 2 and 4. The models' input dimension was defined by the corresponding input feature's dimension, whilst the auxiliary label dimension was established as 3 for SEED and 5 for SEED-V. The cWGAN discriminator output dimension was set at 1, whereas for the cBEGAN discriminator, the output dimension corresponded with its input dimension. 
A linear SVM was used with cross-validation conducted on both datasets for robust evaluation. Notably, cBEGAN exhibited superior convergence performance compared to cWGAN. By measuring the discrepancy between the reconstruction loss distributions rather than the data distributions themselves, cBEGAN effectively captured more intricate information about the real data distribution, enabling the generation of artificial samples with rich information and diverse characteristics. 
When applying cBEGAN on the SEED dataset, authors observed a maximum mean accuracy of 87.56\% when 2,000 samples were added (a 5.66\% increase from the baseline). On the other hand, a maximum of 2.02\% increase was observed for using cWGAN, that was when 200 samples was added. For SEED-V dataset, maximum mean accuracy was 62.87\% when 2,000 samples were generated using cBEGAN, an 8.53\% increase from the baseline; whereas using cWGAN only resulted in a 5.31\% maximum increase.

\section{Discussion}\label{sec:discussion}

With the challenges in building a reliable neurophysiological emotion recognition system, such as the limited number of large labeled datasets, low signal-to-noise (SNR) ratio, and  costly, labor-intensive data collecting process in HCI~\cite{sSaha2021,bAri2022}, applying generative models to increase sample volume has yielded promising results in neuronal emotion recognition frameworks. The ability of VAEs and GANs to capture complex patterns in neuronal data and generate realistic synthetic samples has been demonstrated across multiple studies in this review. These models have been successful in addressing the limited availability of original datasets, thus expanding the potential for training emotion recognition systems with more representative data.

\subsection{Trends of Affective Generative Frameworks in Literature}

\textit{Input Formulation:} Considering the complexity of the affective neurophysiological data and unavoidable artifacts, input formulation is vital in any generative framework. An important observation from the analyzed publications is the exclusive focus on augmenting EEG signals using generative models in the context of neuronal emotion recognition. There is a notable absence of studies exploring the application of generative models to synthesize affective fNIRS signals. This may be attributed to the lack of publicly available multimodal affective databases containing fNIRS signals. Other than \cite{Zhang2022_GANSER}, which proposed their model in data space, the rest of the reviewed paper used feature space as input to synthesize and classify their data. Notably, PSD and DE are the most frequently used features and have provided promising results in different emotion recognition systems. Furthermore, raw neurophysiological signals often lack an inherent representation of the connections between brain regions. To address this lack of spatial information, some studies ~\cite{gBao2021} transformed the 1D EEG channel set to a 2D mesh-like representation by  mapping  recorded data to the positions of the electrodes based on predefined international montages. Additionally, to reduce the impact of noise and due to the rapid temporal nature of emotions, most of the existing framework segments the EEG signals into smaller chunks between 1 and 10 seconds before extracting PSD or DE features. Segmentation reduces non-stationary characteristics in the signal, thereby enabling the generative model to better understand the underlying distribution and capture more localized dynamics. It is worth noting that this segmentation approach already increased the training size, contributing to a more accurate classification model~\cite{yPeng2022}. Generally, feature extraction, spatial information, and segmentation were common formulas in most of the reviewed papers. 

\textit{Model Deployment:} In terms of the use of generative models, most of the existing frameworks predominantly favor the use of wGAN-GP. Despite the stability of VAE training and its inherent method of not overfitting, concerns over the quality of generated samples and strong assumptions about the data held in VAE have pushed researchers towards GAN-based models. The typical implementation of wGAN-GP often incorporates the emotion label as an auxiliary condition to guide the model in learning the hidden distribution of each class. Furthermore, only one study~\cite{dBethge2022} considered individual differences and conditioned their framework to different subjects. 

\textit{Evaluation:} In this review, we observed two strategies in using the augmented data: selective-usage and full-usage~\cite{yLuo2020}, with a focus on the quality and quantity of generated samples used to supplement the training set. While some authors directly appended all synthesized data to the training set, others such as Zhang et al.~\cite{aZhang2021}. and Luo et al.~\cite{yLuo2020} after evaluating the quality of the generated data, used part of the synthesized data for training. While various evaluation methods were employed, including discriminator loss, MMD, t-SNE visualization, and classification accuracy, there is no consensus regarding which evaluation method is most representative of data quality. Additionally, most of the studies investigated the volume of generated samples for augmentation; however, no common pattern was observed related to the quantity of the synthesized samples. Luo et al.~\cite{yLuo2020} suggested that the optimal performance is achieved when the augmented dataset is less than ten times the original size, but this may be influenced by aspects like the generative model, dataset features, and classifier choice. Furthermore, most of the reviewed papers evaluated their framework in subject-dependent or subject-biased experiments.

\subsection{Future Directions in Affective Generative Framework}
In this section, we have provided the aspects that have been missing in the investigating chain of the generative frameworks for affective neurophysiological data augmentation. 

Current generative frameworks primarily concentrate on feature space for synthesizing distinct brain signal features. These features, such as DE and PSD, are derived from deterministic functions. It may be worthwhile to investigate various non-parametric methods (without any assumption about the data) aimed at estimating robust abstractions of affect-relevant multi-channel neurophysiological signals, such as EEG. Furthermore, exploring different segmentation techniques, such as those employing mutual information~\cite{lPiho2018}, and investigating the efficacy of utilizing complete trials against employing segmented data could present a valuable opportunity to enhance the accuracy of affective signal synthesis. An analytic understanding of the respective merits and potential drawbacks of these divergent approaches could yield valuable insights into the optimal affective neurophysiological representation for improving the performance of generative augmentation frameworks.

Acknowledging the complexity of emotions, individual perception variations, and the need for diverse data for a robust emotion recognition system, it would be beneficial to incorporate subjective differences such as age, gender, and cognitive state, as well as emotional categories, into the augmentation framework. Understanding how these different subjective factors affect the performance of generative models can be critical in creating more accurate and personalized emotion recognition systems. This approach, akin to the one presented in ~\cite{dBethge2022}, could open up valuable avenues for constructing a comprehensive end-to-end emotion recognition system trained on a sufficient variety of subjects. Moreover, the majority of the proposed frameworks have been tested primarily on subjective or subject-biased experiments. Given the initial rationale for using generative frameworks in emotion recognition systems - that being to supply meaningful diversity - testing these frameworks on subject-independent experiments could further refine the understanding of the essential elements of the augmentation framework that may need modification or replacement.

The evaluation metrics for the quality of generated data and the necessary volume of augmented data to append to the training dataset currently lack consensus. This underscores the necessity for additional research and standardization in assessing the quality of synthesized neurophysiological data. Stepping toward this goal can endorse reproducibility in other studies. In terms of classifier selection, both SVM and DNN have been prevalently used in the literature reviewed. As observed in ~\cite{aZhang2021}, complexity in the synthesizing framework does not invariably lead to enhanced classification accuracy, and certain classifiers may perform more superior with simpler frameworks. Future research could consider investigating the role and relation of various classifiers with different augmentation techniques. Such insights can assist in identifying the optimal pathway between the classifier and augmentation model, leading to improved accuracy and overall performance in neurophysiological emotion recognition systems. 

Finally, given the limited yet different diversity of neurophysiological signals collected in public affective databases, there's an opportunity to create generative models that can augment across different physiological modalities. For instance, learning the mapping between EEG and fNIRS signals could potentially allow for the augmentation of one type of signal from the other. This cross-modal augmentation could be useful in situations where only a subset of signal types is available or for bridging between datasets collected using different recording techniques. Furthermore, this approach could provide a richer and more holistic representation of emotional states, as it would integrate information from multiple physiological sources. 

\section{Conclusion}\label{sec:conclusion}
As we increasingly integrate technology into every aspect of our daily lives, the demand for machines capable of understanding and responding to human emotions has surged dramatically. This compelling need calls for the development of reliable end-to-end emotion recognition systems, capable of enhancing human-computer interaction by bridging the gap between human affective states and machine perception. While neurophysiology is a promising avenue for constructing sophisticated emotion recognition systems, a notable bottleneck hindering this advancement is the limitations of large public affective datasets containing neurophysiological data, rendering the task of training competent models challenging. 

In this paper, we reviewed existing generative frameworks that address the data scarcity issue in neuronal emotion recognition systems. We delved into various aspects, including input formulation, model deployment, and the necessity for establishing standard evaluation metrics in the physiological data space that require further investigation. Additionally, we highlight the importance of exploring cross-modality and cross-subject augmentation to gain a better understanding of the underlying dynamics of the brain. By doing so, we can build a reliable and accurate end-to-end emotion recognition system that accounts for the complexity of emotions.

\bibliographystyle{elsarticle-num-names} 
\bibliography{main}





\end{document}